\documentclass[aps,preprint]{revtex4}%
\usepackage{amsfonts}
\usepackage{amsmath}
\usepackage{amssymb}
\usepackage{graphicx}%
\begin{document}
\title{Spin Current Through a Magnetic-Oscillating Quantum Dot}
\author{Ping Zhang$^{1,2}$, Qi-Kun Xue$^{1}$, and X.C. Xie$^{1,2}$}
\address{$^{1}$International Center for Quantum Structures, Institute of Physics,
Chinese Academy of Sciences, Beijing 100080, P.R. China\\
$^{2}$Department of Physics, Oklahoma State University, Stillwater, OK 74075}

\begin{abstract}
Non-equilibrium spin transport through an interacting quantum dot is analyzed.
The coherent spin oscillations in the dot provide a generating source for spin
current. In the interacting regime, the Kondo effect is influenced in a
significant way by the presence of the precessing magnetic field. In
particular, when the precession frequency is tuned to resonance between spin
up and spin down states of the dot, Kondo singularity for each spin splits
into a superposition of two resonance peaks. The Kondo-type cotunneling
contribution is manifested by a large enhancement of the pumped spin current
in the strong coupling, low temperature regime.%

\end{abstract}
\maketitle

\textit{Introduction}.---The rapid progress of nano-electronics and
information technologies has prompted intense interest in exploiting the
interplay of electron charge and spin properties, which results in the
emergence of semiconductor spintronics\cite{Datta,Prinz}. Since it is a
manageable task to manipulate and measure electron and nuclear spins via
optical or electric field\cite{Kane}, the solid-state implementation of a
quantum information processing architecture is intensely studied\cite{Samarth}%
, aiming at the possibility of large-scalability of the quantum computer. One
of the most important spin-based electronic devices is a mesoscopic quantum
dot (QD) system, in which the spin coherence time for electrons or nuclei is
relatively long\cite{Kikkawa,Kou}. Spin-polarized transport through a QD has
been extensively investigated recently. It has been shown
theoretically\cite{Mucciolo,Wang} and demonstrated experimentally\cite{Watson}
that a QD system will function as a phase-coherent spin pump in the presence
of sizable Zeeman splitting. A QD-based spin battery device has been proposed
by use of a combination of a vertical magnetic field and an oscillating
electric field\cite{Sun1}. Very recently, spin-polarized current has been
detected from a quantum point contact (QPC)\cite{Potok1} and from
Coulomb-blockaded QDs\cite{Potok2}. For the latter, due to the large charging
energy at low temperatures, a more subtle effect---the creation of new states
of many-body character at the Fermi level by the Kondo effect---is expected to
occur\cite{Hewson}. Although the effect of Kondo resonance on the charge
current has been well studied in a QD\cite{Kouwenhoven}, its influence on the
spin current and spin detection might be equally significant, since it
provides a spin-flip cotunneling mechanism.

In this paper we explore spin current response to a magnetic-oscillating
quantum dot, which is schematically shown in Fig. 1. The setup is similar to
the one used in recent ESR-STM experiments\cite{Wach}. The single-electron
levels in the dot are split by an external magnetic field $B_{0}$,
$\varepsilon_{d\downarrow}-\varepsilon_{d\uparrow}=g\mu_{B}B_{0}$, where $g$
is the effective electron gyromagnetic factor and $\mu_{B}$ is the Bohr
magneton. The two spin levels are coupled by a rotating magnetic field
($B_{1}\cos\omega t$, $B_{1}\sin\omega t$), applied perpendicularly to the
field $B_{0}$. The possibility of the ESR setup based on GaAs QD has been
discussed in Ref.[9]. The QD is in some sense similar to an oscillating
magnetic dipole moment of a two-level atom. However, we show that the spin
transport properties are highly non-trivial for the QD system, due to the
many-body excitations and coupling with an external electrode.

\textit{Theoretical formalism}.---The model Hamiltonian under consideration
can be written as%

\begin{align}
H=\sum_{\sigma}\varepsilon_{d\sigma}d_{\sigma}^{+}d_{\sigma}+Ud_{\uparrow}%
^{+}d_{\uparrow}d_{\downarrow}^{+}d_{\downarrow}  &  -g\mu_{B}B_{1}%
(d_{\uparrow}^{+}d_{\downarrow}e^{i\omega t}+d_{\downarrow}^{+}d_{\uparrow
}e^{-i\omega t})\tag{1}\\
+\sum_{k\sigma}\epsilon_{k}a_{k\sigma}^{+}a_{k\sigma}  &  +\sum_{k\sigma
}\left[  V_{k}a_{k\alpha\sigma}^{+}d_{\sigma}+\text{H.c.}\right]  .\nonumber
\end{align}
Here $d_{\sigma}^{+}$ and $a_{k\sigma}^{+}$ create an electron of spin
$\sigma$ in the QD and in the lead, respectively. The first two terms in (1)
represent correlated spin levels of the QD, the time-dependent terms describe
the coupling between two QD spin states $|\uparrow\rangle$ and $|\downarrow
\rangle$ by the rotating magnetic field, and the last term is the
hybridization of the QD to the lead. A similar model has been proposed for
electrical detection of single-electron spin resonance\cite{Engel,Jiang}.

The time dependence of the Hamiltonian (1) may be eliminated by introducing a
unitary transformation $U=e^{-i\frac{\omega t}{2}[\sum_{k}(a_{k\downarrow}%
^{+}a_{k\downarrow}-a_{k\uparrow}^{+}a_{k\uparrow})+(d_{\downarrow}%
^{+}d_{\downarrow}-d_{\uparrow}^{+}d_{\uparrow})]}$ ($\hslash$ is set to be
unity) and thus redefining the Hamiltonian in the rotating reference (RF) as
follows
\begin{align}
H_{RF}  &  =U^{-1}HU+i\frac{dU^{-1}}{dt}U\tag{2}\\
&  =\sum_{\sigma}\widetilde{\varepsilon}_{d\sigma}d_{\sigma}^{+}d_{\sigma
}+Ud_{\uparrow}^{+}d_{\uparrow}d_{\downarrow}^{+}d_{\downarrow}-g\mu_{B}%
B_{1}(d_{\uparrow}^{+}d_{\downarrow}+d_{\downarrow}^{+}d_{\uparrow
})\nonumber\\
&  +\sum_{k\sigma}\epsilon_{k\sigma}a_{k\sigma}^{+}a_{k\sigma}+\sum_{k\sigma
}\left[  V_{k}a_{k\alpha\sigma}^{+}d_{\sigma}+\text{H.c.}\right]  ,\nonumber
\end{align}
where $\widetilde{\varepsilon}_{d\sigma}=\varepsilon_{d\sigma}\pm\frac{\omega
}{2}$ and $\epsilon_{k\sigma}=\epsilon_{k}\pm\frac{\omega}{2}$ are the dressed
dot and lead levels for up and down spins, respectively. One can see that the
rotating magnetic field, in effect, shifts lead electron energy to the
opposite directions for up and down spins. This spin-oriented energy splitting
originates from the hybridization term in (2), which propagates energy from
the QD to the lead. Thus the spin-up chemical potential is now $\mu_{\uparrow
}=-\omega/2$ (The reference energy is set $E_{F}=0$), while the spin-down
chemical potential is $\mu_{\downarrow}=\omega/2$. The charge chemical
potential is still $\mu_{e}=(\mu_{\uparrow}+\mu_{\downarrow})/2=0$. The
distinction between $\mu_{\sigma}$ and $\mu_{e}$ was recently discussed in Ref.[19].

The physical picture of spin current generation is as follows. Due to the
one-photon difference between the spin-down and spin-up chemical potentials of
the lead, an electron in the spin-down channel can tunnel into the
$\widetilde{\varepsilon}_{d\downarrow}$ level in QD. After a spin-flip process
given by $B_{1}$ term in (2), the electron can tunnel out of the QD into the
spin-up channel. This process repeats, so a steady spin current builds up in
the lead. Note that the net charge current is zero because there is only one
electrode in the system. This zero charge current feature is retained when
more electrodes are taken into account with no bias voltage applied between
them. We emphasize that the present situation is not a simple realization of
magnetization of the electrode by the rotating magnetic field. Here the
generation of spin current is due to real charge transfer processes between
the two spin channels of the lead. Therefore, the spin current can be
conveniently controlled in experiments. To calculate the spin current we make
use of the non-equilibrium Green function formalism.

In the absence of hybridization and Coulomb interaction, the QD is reduced to
a simple driven two-level system (TLS). Its dynamics is featured by a coherent
weight transfer (Rabi oscillations) between the two spin states, which is
complete when the rotating frequency is tuned to the resonant condition
$\omega_{R}=g\mu_{B}B_{0}$. The spin oscillation period is $T=\pi/\Omega$,
where $\Omega=\sqrt{\Delta^{2}+4(g\mu_{B}B_{1})^{2}}$ is the Rabi frequency
and $\Delta=\omega_{R}-\omega$ denotes the detuning from the resonance.

To proceed we introduce a canonical transformation
\begin{equation}
\left(
\begin{array}
[c]{l}%
d_{\uparrow}\\
d_{\downarrow}%
\end{array}
\right)  =\mathbf{u}\left(
\begin{array}
[c]{l}%
c_{\uparrow}\\
c_{\downarrow}%
\end{array}
\right)  \text{ with }\mathbf{u}=\left(
\begin{array}
[c]{ll}%
\cos\phi & -\sin\phi\\
\sin\phi & \cos\phi
\end{array}
\right)  , \tag{3}%
\end{equation}
where $\phi=\tan^{-1}(\frac{2g\mu_{B}B_{1}}{\Omega-\Delta})$. In terms of new
fermion operators, the dot Hamiltonian in Eq. (2) is has a diagonal form
$\sum_{\sigma}\varepsilon_{c\sigma}c_{\sigma}^{+}c_{\sigma}+Uc_{\uparrow}%
^{+}c_{\uparrow}c_{\downarrow}^{+}c_{\downarrow}$ with $\varepsilon_{c\sigma
}=(\varepsilon_{d\uparrow}+\varepsilon_{d\downarrow}\pm\Omega)/2$ for the up
and down spins.

The spin-$\sigma$, say spin-$\uparrow$, current can be calculated from the
time evolution of the occupation number $N_{\uparrow}(t)=\sum_{k}c_{k\uparrow
}^{+}c_{k\uparrow}$ for electrons in the electrode using the non-equilibrium
Green function:
\begin{equation}
J_{\uparrow}=-e\frac{d}{dt}\langle N_{\uparrow}(t)\rangle=\frac{2e}{\hslash
}\operatorname{Re}\left\{  \sum_{k}V_{k}\int\frac{d\epsilon}{2\pi}%
[\mathbf{uG}_{k,c}^{<}(\epsilon)]_{\uparrow\uparrow}\right\}  , \tag{4}%
\end{equation}
where we have defined the lesser Green function matrix $[\mathbf{G}_{k,c}%
^{<}(t)]_{\sigma\sigma^{\prime}}=i\langle c_{\sigma^{\prime}}^{+}%
(0)a_{k\sigma}(t)\rangle$. Next, we use Dyson's equation to calculate
$\mathbf{G}_{k,c}^{<}$ and express $J_{\uparrow}$ by the Green functions of
the dot as follows
\begin{equation}
J_{\uparrow}=\frac{ie}{\hslash}\int\frac{d\epsilon}{2\pi}\Gamma(\epsilon
)\left(  \mathbf{u}\{\mathbf{G}_{c}^{<}(\epsilon)+f_{\uparrow}(\epsilon
)[\mathbf{G}_{c}^{r}(\epsilon)-\mathbf{G}_{c}^{a}(\epsilon)]\}\mathbf{u}%
^{\text{T}}\right)  _{\uparrow\uparrow}, \tag{5}%
\end{equation}
where $\Gamma(\epsilon)=2\pi\sum_{k}|V_{k}|^{2}\delta(\epsilon-\epsilon_{k})$
and $[\mathbf{G}_{c}^{<}(t)]_{\sigma\sigma^{\prime}}=i\langle c_{\sigma
^{\prime}}^{+}(0)c_{\sigma}(t)\rangle$ are the lesser Green functions of the
dot, $\mathbf{A}=\mathbf{G}_{c}^{r}-\mathbf{G}_{c}^{a}$ is the spectral
function matrix, $\ $and $f_{\sigma}(\epsilon)=f(\epsilon-\mu_{\sigma})$ are
the Fermi distribution functions for the spin-up and spin-down channels.

\textit{The non-interacting} \textit{spin current}.---In the non-interacting
($U=0$) case or within the mean field (MF) treatment of the Coulomb
interaction for finite $U$, the retarded Green function matrix is given by
\begin{equation}
\lbrack\mathbf{G}_{c}^{r}(\epsilon)]_{\sigma\sigma^{\prime}}=\delta
_{\sigma\sigma^{\prime}}\frac{1}{[\mathbf{g}_{c}^{r-1}(\epsilon)]_{\sigma
\sigma}-[\mathbf{\Sigma}_{0}^{r}(\epsilon)]_{\sigma\sigma}}, \tag{6}%
\end{equation}
where $\mathbf{\Sigma}_{0}^{r}=-i\Gamma\mathbf{I}$ is the tunneling
contribution to the retarded self-energy, and $\mathbf{g}_{c}^{r}(\epsilon)$
is the unperturbed Green function matrix of the dot, which in the mean-field
approximation has a simple Hartree-Fock type $[\mathbf{g}^{R}(\epsilon
)]_{\sigma\sigma^{\prime}}=\delta_{\sigma\sigma^{\prime}}[\frac{n_{c\overline
{\sigma}}}{\epsilon-\epsilon_{c\sigma}-U}+\frac{1-n_{c\overline{\sigma}}%
}{\epsilon-\epsilon_{c\overline{\sigma}}}]$ with $n_{c\sigma}=\langle
c_{\sigma}^{+}c_{\sigma}\rangle$. $n_{c\sigma}$ must be calculated
self-consistently via the relation $n_{c\sigma}=\operatorname{Im}\int
\frac{d\epsilon}{2\pi}[\mathbf{G}_{c}^{<}(\epsilon)]_{\sigma\sigma}$. The
lesser Green functions in (5) are obtained via the Keldysh equation
$\mathbf{G}_{c}^{<}=\mathbf{G}_{c}^{r}\mathbf{\Sigma}_{0}^{<}\mathbf{G}%
_{c}^{a}$ with the lesser self-energy matrix
\begin{equation}
\mathbf{\Sigma}_{0}^{<}=i\Gamma\left(
\begin{array}
[c]{ll}%
f_{\uparrow}(\epsilon)\cos^{2}\phi+f_{\downarrow}(\epsilon)\sin^{2}\phi &
\sin\phi\cos\phi\lbrack f_{\downarrow}(\epsilon)-f_{\uparrow}(\epsilon)]\\
\sin\phi\cos\phi\lbrack f_{\downarrow}(\epsilon)-f_{\uparrow}(\epsilon)] &
f_{\uparrow}(\epsilon)\sin^{2}\phi+f_{\downarrow}(\epsilon)\cos^{2}\phi
\end{array}
\right)  . \tag{7}%
\end{equation}
Substituting the expressions for $\mathbf{G}_{c}^{<}$ into (5), we obtain the
expression for the tunneling spin current
\begin{equation}
J_{\uparrow}=\frac{e\Gamma}{\hslash}\int\frac{d\epsilon}{2\pi}\{\mathbf{uG}%
_{0}^{r}\mathbf{(\epsilon)}\overline{\mathbf{\Sigma}}_{0}^{<}\mathbf{G}%
_{0}^{a}\mathbf{(\epsilon)u}^{T}\}_{\uparrow\uparrow}\{f_{\uparrow}%
(\epsilon)-f_{\downarrow}(\epsilon)\}, \tag{8}%
\end{equation}
where
\begin{equation}
\overline{\mathbf{\Sigma}}_{0}^{<}=\Gamma\left(
\begin{array}
[c]{ll}%
\sin^{2}\phi & \sin\phi\cos\phi\\
\sin\phi\cos\phi & \cos^{2}\phi
\end{array}
\right)  . \tag{9}%
\end{equation}

Since the electrons are flowing from spin-down channel to spin-up channel, we
define the total spin current as $J_{s}=J_{\downarrow}-J_{\uparrow
}=-2J_{\uparrow}$. Although the coherent spin superposition is strongest at
two-level resonance $\omega=\omega_{R}$, the spin current is also influenced
by the spin levels $\widetilde{\varepsilon}_{d\sigma}$relative to $\mu
_{\sigma}$. Figure 2(a) shows the non-interacting spin current versus rotating
frequency at different temperatures. The undressed spin levels are set to lie
deeply below $E_{F}$. One can see that at low temperature (solid line), the
spin current peak is not at, but far from $\omega=\omega_{R}$. This is because
at $\omega=\omega_{R}$ the dressed spin-up level $\widetilde{\varepsilon
}_{d\uparrow}$ is still lower than $\mu_{\uparrow}$ and thus the electron in
the dot cannot tunnel out via small thermal excitation. Due to the state
exchange in QD, as shown in the inset of Fig. 2(a), when $\omega$ increases
further, crossing the avoided crossing between the two spin levels, the upper
eigenstate is dominated by spin component $|\uparrow\rangle$ while the lower
state by $|\downarrow\rangle$. Thus when the upper level is higher than
$\mu_{\uparrow}$, spin tunneling processes occurs, and a peak develops in the
current spectrum. When the temperature is increased to $k_{B}T\sim
|\mu_{\uparrow}-\widetilde{\varepsilon}_{d\uparrow}|$, the spin-current peak
shifts to $\omega=\omega_{R}$ with a larger peak amplitude. Further increasing
temperature such that $k_{B}T>\omega$ will smear out the mismatch between
$\mu_{\uparrow}$ and $\mu_{\downarrow}$, and thus the spin current amplitude
begins to decrease, as shown in Fig. 2(a). The dot levels can be conveniently
controlled by the gate voltage $V_{g}$. When $V_{g}$ is modulated such that
$\widetilde{\varepsilon}_{d\sigma}$ are at the interval between $\mu
_{\downarrow}$ and $\mu_{\uparrow}$, then the spin current peak will occur
exactly at $\omega=\omega_{R}$ [see Fig. 2(b)].

The model can also illustrate the spin-current generation in the Coulomb
blockade regime when Coulomb interaction $U$ is treated in the mean-field
formalism. As an example, we show in Fig. 3 the spin current versus gate
voltage for several values of $\omega$. Two Coulomb peaks with interval $U$
can be resolved, and as in Fig. 2, the current amplitude increases when the
magnetic field frequency is tuned towards to two-level resonance
$\omega=\omega_{R}$.

\textit{The} \textit{electronic} \textit{correlation effect}.---In the low
temperature, strong coupling, and large charging energy regime, correlation
effects enter and are expected to significantly influence the spin transport.
To illustrate correlation effects, we again use the equation of motion method
to solve the retarded Green function in (2), which generates higher-order
Green functions that have to be truncated to close the equation\cite{Meir}.
After a straightforward calculation, in the infinite-$U$ limit we obtain
\begin{equation}
\mathbf{G}^{r}(\epsilon)=[\epsilon\mathbf{I}-\widehat{\widetilde
{\mathbf{\varepsilon}}}_{d}-\mathbf{\Sigma}_{0}^{r}(\epsilon)-\mathbf{\Sigma
}_{1}^{r}(\epsilon)]^{-1}(\mathbf{I}-\mathbf{n}_{c}), \tag{10}%
\end{equation}
where $(\widehat{\widetilde{\mathbf{\varepsilon}}}_{d})_{\sigma\sigma^{\prime
}}=\delta_{\sigma\sigma^{\prime}}\widetilde{\varepsilon}_{d\sigma}$ and the
interaction self-energies are given by $[\mathbf{\Sigma}_{1}^{r}%
(\epsilon)]_{\sigma\sigma^{\prime}}=\delta_{\sigma\sigma^{\prime}}\sum
_{k}[\frac{\sin^{2}\phi|V_{k}|^{2}f(\epsilon_{k\sigma})}{\epsilon
-\epsilon_{k\sigma}-\varepsilon_{c\sigma}+\varepsilon_{c\overline{\sigma}}%
}+\frac{\cos^{2}\phi|V_{k}|^{2}f(\epsilon_{k\overline{\sigma}})}%
{\epsilon-\epsilon_{k\overline{\sigma}}-\varepsilon_{c\sigma}+\varepsilon
_{c\overline{\sigma}}}]$. $(\mathbf{n}_{c})_{\sigma\sigma^{\prime}}%
=\delta_{\sigma\sigma^{\prime}}\langle c_{\overline{\sigma}}^{+}%
c_{\overline{\sigma}}\rangle$ needs to be calculated self-consistently via the
relation $\langle c_{\sigma}^{+}c_{\sigma}\rangle=\operatorname{Im}\int
\frac{d\epsilon}{2\pi}\mathbf{G}_{\sigma\sigma}^{<}(\epsilon)$. $\mathbf{G}%
^{<}(\epsilon)$ is difficult to obtain since the lesser self-energy cannot be
given exactly in the interacting regime. However, the situation simplifies in
the steady-state transport\cite{Sun2}, in which the occupation on the dot is
time-independent, $i\frac{d}{dt}\langle c_{\sigma}^{+}(t)c_{\sigma}%
(t)\rangle=0$. Writing down the Heisenberg equations for $c_{\sigma}^{+}(t)$
and $c_{\sigma}(t)$, and integrating out the lead electron operators, we find
the self-consistent integral equations for the retarded Green functions as
follows:
\begin{equation}
\langle c_{\uparrow}^{+}c_{\uparrow}\rangle=\operatorname{Im}\int
\frac{d\epsilon}{2\pi}\mathbf{G}_{\uparrow\uparrow}^{<}(\epsilon)=\int
\frac{d\epsilon}{2\pi}\operatorname{Im}(\mathbf{G}_{\uparrow\uparrow}%
^{a}-\mathbf{G}_{\uparrow\uparrow}^{r})[\cos^{2}\phi f_{\uparrow}+\sin^{2}\phi
f_{\downarrow}], \tag{11a}%
\end{equation}%
\begin{equation}
\langle c_{\downarrow}^{+}c_{\downarrow}\rangle=\operatorname{Im}\int
\frac{d\epsilon}{2\pi}\mathbf{G}_{\downarrow\downarrow}^{<}(\epsilon
)=\int\frac{d\epsilon}{2\pi}\operatorname{Im}(\mathbf{G}_{\downarrow
\downarrow}^{a}-\mathbf{G}_{\downarrow\downarrow}^{r})[\sin^{2}\phi
f_{\uparrow}(\epsilon)+\cos^{2}\phi f_{\downarrow}(\epsilon)], \tag{11b}%
\end{equation}
which now, together with (10), close the equations of motion for
$\mathbf{G}^{r}$. Similarly, the quantities $\langle c_{\sigma}^{+}%
c_{\overline{\sigma}}\rangle=-i\int\frac{d\epsilon}{2\pi}G_{\sigma
\overline{\sigma}}^{<}(\epsilon)$ can also be expressed by the retarded Green
functions. Thus we can calculate the spin current in (5) directly after
$\mathbf{G}^{r}(\epsilon)$ is numerically obtained.

The magnetic-field-induced excitation properties are shown in Fig. 4 by
plotting local spin-resolved spectral densities $\rho_{\sigma}(\epsilon
)=[\mathbf{uG}_{c}^{r}(\epsilon)\mathbf{u}^{\text{T}}]_{\sigma\sigma}$ for
different values of $\omega$. In the absence of a rotating magnetic field
($\omega,B_{1}=0$), as observed from Fig. 4(a), the spectral density for each
spin component is characterized by a broad single-particle peak around
$\varepsilon_{d\sigma}$, and a sharp Kondo peak at Zeeman energies
$\epsilon=-\omega_{R}$ for up spin and $\epsilon=\omega_{R}$ for down spin.
The spectral weight of $\rho_{\uparrow}$ is enhanced by the magnetic field,
while $\rho_{\downarrow}$ is greatly suppressed. Therefore, a net spin moment
develops in the QD. Physically, the Kondo peak at $\epsilon=-\omega_{R}$ is
due to the cotunneling processes (see inset) in which one spin-down electron
in the lead at the Fermi level tunnels into the dot and occupies energy level
$\widetilde{\varepsilon}_{\downarrow}$, followed by another electron on
$\widetilde{\varepsilon}_{\uparrow}$ tunneling out to the spin-up channel with
the energy $\omega_{R}$ below from Fermi level $E_{F}$. Such cotunneling
processes transfer electrons from spin-up channel to spin-down channel of the
lead. Thus\textit{\ a spin current may be induced in an efficient way by these
Kondo-type cotunneling processes}. Similarly, the Kondo peak at $\epsilon
=\omega_{R}$ in $\rho_{\downarrow}(\epsilon)$ is due to a superposition of
many cotunneling processes in which one spin down electron with the energy
$\omega_{R}$ above the Fermi level tunnels into the dot and simultaneously
another spin-up electron in the dot tunnels out to the Fermi level of the
lead. Both kinds of cotunneling processes contribute to the spin-current
generation. When the rotating magnetic field is switched on, the spin Kondo
resonances and spectral structures begin to change. At $\omega<\omega_{R}$,
the broad single-spin and sharp Kondo peaks shift toward higher energies for
up spin, while they shift to lower energies for down spin [Fig. 4(b)]. At
two-level resonance $\omega=\omega_{R}$, as shown in Fig. 4(c): (i) each spin
Kondo singularity splits into two prominent peaks around $\mu_{\uparrow}$ and
$\mu_{\downarrow}$; (ii) $\rho_{\uparrow}(\epsilon)$ and $\rho_{\downarrow
}(\epsilon)$ completely overlap with identical spectral weight. This overlap
is caused by the fact that at $\omega=\omega_{R}$, each spin eigenstate in the
dot is a strong superposition of $|\uparrow\rangle$ and $|\downarrow\rangle$
states with equal probability amplitude. When $\omega$ increases further, the
two-level superposition is suppressed, and the state exchange occurs. As a
response, the spectral densities of two spins also exchange their structures;
as can be seen from Fig. 4(d), $\rho_{\downarrow}(\epsilon)$ is enhanced while
$\rho_{\uparrow}(\epsilon)$ is greatly suppressed.

To see the influence of the many-body correlation effect on the spin
transport, we show in Figure 5 the spin current as a function of rotating
frequency. Compared to the non-interacting case, it can be seen that the spin
current is greatly enhanced by the cotunneling processes in the strong
coupling, low temperature regime. A detailed study of cotunneling effects on
the spin current is in progress.

In summary, we have analyzed the spin current properties of an interacting QD
exposed to a rotating magnetic field. The spin-flip process in the QD and the
effective difference of two spin chemical potentials suffices the generation
of spin current with no charge current. The interplay of the many-body
cotunneling process and coherent two-level resonance in the QD produces a
remarkable enhancement of spin current. Since a controllable generation of
spin current provides an efficient source for spin injection, spin detection,
and virous variety of spin-based devices, we expect the present results may
have practical applications in the field of spintronics.

This work was supported by grants CNSF 90103024, MOST-G001CB3095, and U.S.-DOE-DE-FG03-98ER45687.

{\Large Figure captions}

Fig. 1. Model setup for spin transport through a QD. The degenerate spin
levels in the dot are split and coupled by the external magnetic fields
$B_{0}$ and $B_{1}$, respectively. Due to the imbalance between spin-up and
spin-down chemical potentials (see text), the electron tunnels from the
spin-down channel, through the QD, to the spin-up channel.

Fig. 2. (a) Non-interacting spin current as a function of rotating frequency
at different temperatures with the undressed spin levels set $\varepsilon
_{d\uparrow}=-2.5$, $\varepsilon_{d\downarrow}=-1.5$; (b) Contour plot of spin
current as a function of gate voltage and rotating frequency. The spin levels
in the dot are modulated by the gate voltage as $\varepsilon_{d\sigma}%
(V_{g})=\varepsilon_{d\sigma}+eV_{g}$ with $\varepsilon_{d\uparrow}=1$,
$\varepsilon_{d\downarrow}=2$. Other parameters are $\Gamma=0.1$ and $g\mu
_{B}B_{1}=0.2$.

Fig. 3. Spin current versus gate voltage for different rotating frequencies
with $U=2$. Other parameters are the same as in Fig. 2(b).

Fig. 4. The spin up (solid line) and spin down (dashed line) spectral
densities at different rotating frequency. Parameters are $\varepsilon
_{d\uparrow}=-2.5\Gamma$, $\varepsilon_{d\downarrow}=-1.5\Gamma$,
$T=0.001\Gamma$, and $g\mu_{B}B_{1}=0.2\Gamma$.

Fig. 5. Spin current as a function of rotating frequency in the Kondo-type
cotunneling regime (solid line). The non-interacting spin current is also
shown (dotted line) for comparison. Other parameters are the same as in Fig. 4.

\end{document}